\documentclass[journal=jacsat,manuscript=article,layout=twocolumn]{achemso}

\usepackage[utf8]{inputenc}
\usepackage[english]{babel}
\usepackage{amsmath}
\usepackage{graphicx}
\usepackage{graphics}
\usepackage{epsfig}
\usepackage{caption}
\usepackage{epstopdf}
\usepackage{lipsum}
\usepackage{wrapfig}
\usepackage{lmodern}
\usepackage{cuted}
\usepackage[colorlinks=true, allcolors=blue]{hyperref}
\usepackage{textgreek}
\usepackage[version=3]{mhchem} % 
\DeclareUnicodeCharacter{2003}{ }

\author{Madona Mekhael}
\author{Timo Stolt}
\author{Anna Vesala}
\affiliation[TAU]
{Photonics Laboratory, Physics Unit, Tampere University, FI-33014 Tampere, Finland}
\author{Heikki Rekola}
\affiliation[UEF]
{Center for Photonics Sciences, University of Eastern Finland, P.O. Box 111, FI-80101, Joensuu, Finland}
\author{Tommi K. Hakala}
\affiliation[UEF]
{Center for Photonics Sciences, University of Eastern Finland, P.O. Box 111, FI-80101, Joensuu, Finland}
\author{Robert Fickler}
\affiliation[TAU]
{Photonics Laboratory, Physics Unit, Tampere University, FI-33014 Tampere, Finland}
\author{Mikko J. Huttunen}
\affiliation[TAU]
{Photonics Laboratory, Physics Unit, Tampere University, FI-33014 Tampere, Finland}
\email{mikko.huttunen@tuni.fi}

\title[Phase-Matched Second-Harmonic Generation from Metasurfaces Inside Multipass Cells]
  {Phase-Matched Second-Harmonic Generation from Metasurfaces Inside Multipass Cells}
\abbreviations{IR,NMR,UV}

\begin{document}
\maketitle
%ABSTRACT HERE
    \begin{strip}   % <---
        We demonstrate a simple and scalable approach to increase conversion efficiencies of nonlinear metasurfaces by incorporating them into multipass cells and by letting the pump beam to interact with the metasurfaces multiple times. We experimentally show that by metasurface design, the associated phase-matching criteria can be fulfilled. As a proof of principle, we achieve phase matching of second-harmonic generation (SHG) using a metasurface consisting of aluminium nanoparticles deposited on a glass substrate. The phase-matching condition is verified to be achieved by measuring superlinear dependence of the detected SHG as a function of number of passes. We measure an order of magnitude enhancement in the SHG signal when the incident pump traverses the metasurface up to 9 passes. Results are found to agree well with a simple model developed to estimate the generated SHG signals. We also discuss strategies to further scale-up the nonlinear signal generation. Our approach provides a clear pathway to enhance nonlinear optical responses of metasurface-based devices. The generic nature of our approach holds promise for diverse applications in nonlinear optics and photonics.

        \vspace{0.3cm}      
    \end{strip}    % <---
        \medskip
\section{Introduction}
Nonlinear optical responses of materials are of paramount importance in a wide spectrum of modern photonic applications ranging from the development of ultrafast high-power laser sources~\cite{Ma:19, Fischer:77, Carlson:19}, and optical metrology~\cite{Kasparian:08} to recent realizations of optical neural networks~\cite{Sui:20}. The main challenge in nonlinear optics is often to realize material systems where the inherently weak nonlinearly generated signal fields are, upon propagation through the system, coherently built-up to strengths of practical relevance. In other words, the challenge is to phase match the system~\cite{BoydBook}, which can be achieved e.g.~using special nonlinear crystals~\cite{Zhang:17}, fibers~\cite{Betourne:08}, waveguides~\cite{Dimitropoulos:04}, and/or resonators made of such materials~\cite{Nitiss:22, Hajati:22}. 

Metamaterials are artificial structures made of sub-wavelength building blocks~\cite{Liu:11}. Interestingly, metamaterials can exhibit properties not (easily) achieved using natural materials~\cite{Smith:04,Linden:06},
and have recently emerged as a powerful technology for realizing novel flat photonic components, such as metalenses and meta-holograms~\cite{Aieta:15,Khorasaninejad:16}. 
Nonlinear optical metamaterials have also been recently suggested as a solution for the phase-matching problem, because optically thin nonlinear metamaterials are virtually free of phase-matching issues~\cite{Litch:18}.
%Refs, also a ref of transverse phasematching, Suchowski/Ellenbogen comes first to mind.

Despite advantages and steady progress of nonlinear metasurfaces/metamaterials, their usefulness for applications is still mostly limited by their poor conversion efficiencies~\cite{Vermeulen:23}. Approaches to enhance their nonlinear responses have included various works utilizing resonance enhancement~\cite{Kauranen:12, Stolt:22}, mode-overlap optimization~\cite{Noor:20} and/or index-near-zero behavior~\cite{Alam:16}. Recently, nonlinear responses have also been enhanced by phase-engineered bulk nonlinear metamaterials, consisting of several layers of nonlinear metasurfaces~\cite{Stolt:21}. Despite steady progress, it seems necessary to come up with new approaches to further enhance the nonlinear responses of metasurfaces/materials.  

Here, we experimentally demonstrate an easy approach to enhance nonlinear responses of metamaterials, by incorporating them inside multipass cells and allowing the pump beam to pass through the metamaterial several times. As a proof-of-principle demonstration, we measure second-harmonic generation (SHG) emission from a plasmonic metasurface and demonstrate an order of magnitude enhancement in the measured SHG signal at the wavelengths where the metasurface was phase-matched. Notably, we systematically investigate the dependence of SHG emission on the number of passes, distinctly demonstrating clear signatures of successful phase matching. Importantly, this demonstrated approach is quite generic, offering compatibility with various existing enhancement techniques.

\section{Methods}
\subsection{Semi-analytical model}
%PHASE MATCHING CONDITION
The phase-matching condition of the second-harmonic response within metamaterials can be expressed as
\begin{equation} \label{Eq:mismatch}
\Delta k=2\left(\varphi_\omega+\delta_\omega\right)-\varphi_{2 \omega}-\delta_{2 \omega}=2 \pi m,
\end{equation}
where $m$ is an integer. The terms $\varphi_\omega=k_\omega h$ and $\varphi_{2\omega}=k_{2\omega} h$ denote the phase accumulations of the fundamental and second-harmonic fields, respectively, due to their propagation through a distance $h$. Terms $\delta_\omega$ and $\delta_{2\omega}$ denote the phase shifts incurred due to the interaction between the fundamental and second-harmonic fields with the metasurface, respectively. These phase shifts can be determined from the optical responses of the metamaterials (here from their transmittance spectra), enabling to fulfill the phase-matching condition through appropriate metamaterial design.

By measuring the linear transmission $T$ of the samples, we can determine their phase responses and subsequently estimate the strength of the phase-matched second-harmonic signal upon propagation of the sample metasurfaces multiple times. The transmission $T$ is linked to the polarizability $\alpha$, dictated by the localized surface plasmon resonances (LSPRs) via
\begin{equation} \label{Eq:extinction}
1-T \propto 4 \pi k \operatorname{Im}(\alpha),
\end{equation}
where $k=2\pi/\lambda$ represents the wavevenumber. Assuming $\alpha$ to be a complex-valued Lorentzian function, the full complex-valued $\alpha$ can be found after solving the imaginary part of $\alpha$ from the above equation.
The phase shifts $\delta$ associated with the sample can then be found by using $\alpha=\lvert\alpha\rvert \exp( \mathrm{i} \delta)$.

In the case when the pump beam passes many times through the metasurface, the SHG response can be estimated using the approach already introduced for the stacked metasurfaces \cite{Stolt:21}. The phase-matched SHG signal is given by
\begin{equation} \label{Eq:SHG}
\begin{split}
\mathrm{SHG} &\propto|E(2 \omega)|^2 \\ &\propto\left|\sum_{J=1}^N T(\omega)^J T(2 \omega)^{J / 2} \mathrm{e}^{\mathrm{i} J \Delta k} \chi_{m s}^{(2)} E(\omega)^2\right|^2,
\end{split}
\end{equation}    
where $\chi_{m s}^{(2)}$ is the nonlinear susceptibility of the metasurface and $N$ is the number of passes through the metasurface. The transmittance $T$ near the pump and SHG frequencies is experimentally measured, while the 
phase mismatch $\Delta k$ is calculated using Eq.~\eqref{Eq:mismatch}.

 \begin{figure*}
     \centering
     \includegraphics{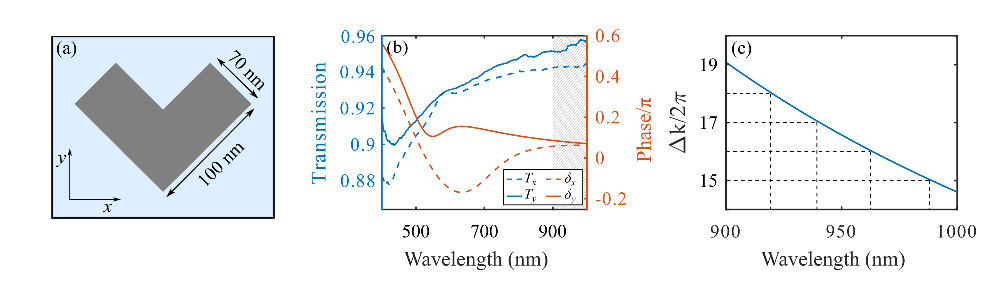}
     \caption{(a) A schematic representing a single V-shaped nanoparticle from the metasurface. 
     (b) The transmission spectrum measured from the sample and the calculated phase changes for $x$- and $y$-polarizations. (c) The calculated phase term $\Delta k$ for the SHG signal emitted from the sample in the wavelength range of the pump (900--1000~nm). Dashed lines highlight the wavelengths (920~nm, 940~nm, 964~nm, and 989~nm) at which the phase-matching condition is fulfilled.}
     \label{fig:spectrum}
 \end{figure*}
 
The sample used in our experiments consisted of V-shaped aluminum nanoparticles (Fig.~\ref{fig:spectrum}a) deposited on a 0.5~mm-thick $\mathrm{SiO_2}$ substrate using standard electron-beam lithography techniques. The nanoparticles had a thickness of 30~nm, arm lengths of 140~nm, and arm widths of 70~nm. The nanoparticles were positioned randomly (being however similarly oriented) with a particle density of 11.11 particles/$\mu \mathrm{m^2}$. This particle density corresponded to density of a periodic square nanoparticle array with periodicity of 300 nm. The reason to investigate randomly positioned nanoparticle arrays was that such arrays do not exhibit collective responses, such as surface lattice resonances~\cite{Kravets2018,Saad2021}, that would unnecessarily complicate the phase extraction protocol and subsequent data analysis. Based on numerical simulations (Ansys/Lumerical FDTD), these particle dimensions resulted in LSPRs near 420~nm for $x$-polarized incident light. We confirmed this by measuring transmission spectra shown in Fig.~\ref{fig:spectrum}b (see the Supplemental material for description of the transmission setup [\textbf{LINK HERE BY THE PUBLISHER}]). Due to the presence of LSPRs, the nanoparticles induce phase-changes (red curve in Fig~\ref{fig:spectrum}b) both at pump and SHG wavelengths, which we then used, with the substrate thickness, to calculate $\Delta k$ shown in Fig.~\ref{fig:spectrum}c. At the wavelengths where $\Delta k$ is an integer of $2\pi$, we expect phase matching and coherent build-up of the SHG signal.

\subsubsection{Nonlinear characterization}
\begin{figure*}[h]
    \centering
    \includegraphics{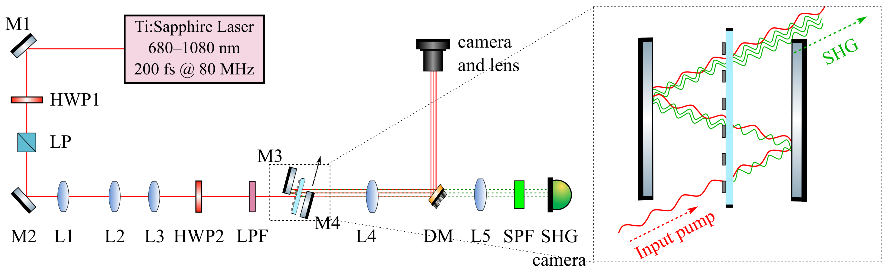}
    \caption{A schematic of the setup used to measure SHG response of the sample at different number of passes $N$. The number of passes is controlled by moving the mirror M4 laterally along the direction shown by the arrow, and observing the shifting of the pump and SHG beams on the cameras. The inset on the right shows the phase-matched SHG emission in case of 3 passes of the pump beam through the metasurface device.
    }
    \label{fig:SHG_setup}
\end{figure*}

The experimental setup, illustrated in Fig.~\ref{fig:SHG_setup}, was used to characterize the SHG response of the sample as a function of the number of passes $N$ of the pump beam through the sample. %It is noteworthy that the generated SHG beam undergoes an equivalent number of passes through the metasurface. 
A tunable titanium sapphire femtosecond laser (Chameleon Vision \MakeUppercase{\romannumeral 2}), operating at an 80 MHz repetition rate and offering a spectral range of 680--1080\;nm with pulse duration of 140\;fs (near 800 nm), was used as the pump source. The FWHM value for the bandwidth of the pump laser at 950 nm was measured to be 5 nm, and consequently a pulse duration of 200 fs was used in our calculations. The combination of the half-wave plate HWP1 and the linear polarizer LP was used to control the power incident on the sample. Furthermore, the half-wave plate HWP2 was used to adjust the polarization of the incident beam to $x$-polarization. Importantly, the use of a low-pass filter LPF before the sample ensured no SHG signal possibly originating from optical components preceding the sample were detected. 

Three lenses, denoted as L1 ($f_1=75$~mm), L2 ($f_2=25.4$~mm), and L3 ($f_3=250$~mm), were used to control the beam size on the sample as well as to adjust the Rayleigh range associated with the Gaussian beam propagation. Using a camera, the beam size before the first lens L1 was measured to be 2.6~mm (FWHM). The lens L2 was positioned to reduce the beam size by a factor of 0.34, resulting in divergence angle of 1.8 mrad for the beam after lens L3. Subsequently, the calculated FWHM value for the beam waist was $343~\mu$m for pump at 950~nm, which enabled to maintain a fairly focused pump beam during its passes through the metasurface (Rayleigh range of 9.75 cm).
We note that a considerably larger Rayleigh range would have resulted in a wider beam profile and correspondingly reduced beam intensity, which subsequently would have resulted in decreased SHG signal. 
 
The number of passes $N$ of the pump beam through the sample was adjusted by displacing mirror M4 along the trajectory indicated by the arrow shown in Fig.~\ref{fig:SHG_setup}. Changes in $N$ resulted in detectable offsets for both the pump and the SHG beams, which were readily recorded with the two cameras used to detect both the pump and the SHG beams. The beam offsets have been highlighted in  Fig.~\ref{fig:SHG_setup} as three lines of varying color intensity. 

The lens L4 ($f_4=50$~mm) was used to collect the generated SH signal emitted from the sample. The dichroic mirror (DM) effectively filtered out the pump beam while reflecting it towards the camera used to image the sample. The SHG beam passed through the DM and was focused by the lens L5 ($f_5=100$~mm) to the SHG camera (ZWO-ASI1600MM). The short-pass filter (SPF) removed any residual pump and made it possible to measure the generated SHG signal.

\section{Results and discussions}
\begin{figure}[ht]
    \centering    \includegraphics{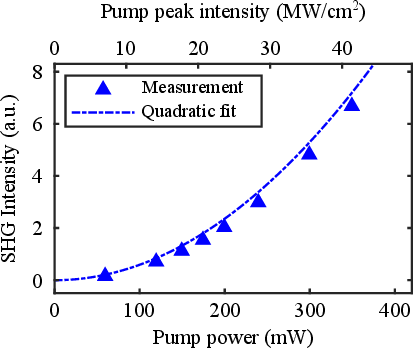}
    \caption{Quadratic dependence of the SHG signal intensity on the  average power and peak power density of the pump laser for one pass through the sample.}
    \label{fig:quadratic}
\end{figure}

First, we measured power dependence of the collected SHG signal for the case of a single pass (Fig.~\ref{fig:quadratic}). The power dependence was verified to be close to quadratic, confirming that the collected signal was of nonlinear origin and that no apparent sample damage occurred at the used power levels. This was not surprising, because we used a relatively large beam waist, and our samples exhibited low absorption at the pump wavelenghts (900--1000~nm).
Consequently, the pump power incident on the metasurface was set to be 350~mW in the following experiments. The $y$-polarized SHG response of the sample was measured as a function of the number of passes for $x$-polarized incident light. 

\begin{figure*}[t]
    \centering
     \includegraphics{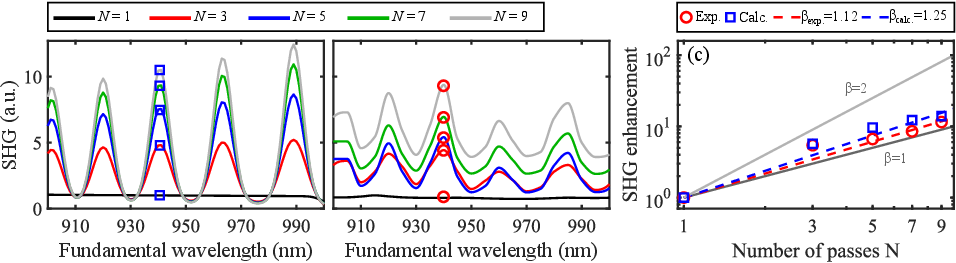}
     \caption{(a) Calculated and (b) measured SHG signal for different number of passes $N$. (c) The enhancement factor of the SHG signal as a function of the number of passes $N$. The calculated ($\beta_\mathrm{pulsed}=1.25$) and the experimentally  measured data ($\beta_\mathrm{exp}=1.12$) are in good agreement and show a superlinear dependence.}
     \label{fig:results}
\end{figure*} 

Our semi-analytical model predicted an enhancement of the SHG response at specific wavelengths as the number of passes $N$ increases, as depicted in Fig.~\ref{fig:results}a. These wavelengths corresponded to the phase-matched wavelengths [see Fig.~\ref{fig:spectrum}c], where the SHG response reached its local maximum. This trend was confirmed by experiments [Fig.~\ref{fig:results}b], where we measured the SHG response as a function of the pump wavelength (900--1000 nm). These measurements were repeated for different number of passes (up to $N=9$ passes). The observed phase-matched wavelengths were found to be 904~nm, 920~nm, 940~nm, 960~nm, and 985~nm, agreeing well with the calculated phase-matched wavelengths [Fig.~\ref{fig:spectrum}c]. 
Notably, the SHG response exhibited a superlinear dependence ($\mathrm{SHG} \propto N^{\beta_\mathrm{exp}}=N^{1.12}$) on the number of passes $N$ of the pump beam, confirming that the SHG signal was successfully phase-matched.

In the ideal scenario of a lossless and perfectly phase-matched sample, the SHG signal would exhibit a quadratic dependence on the number of passes $N$ ($\mathrm{SHG} \propto N^{\beta_\mathrm{ideal}}$, where $\beta_\mathrm{ideal}=2$) \cite{Stolt:21}. However, our sample was neither perfectly lossless nor perfectly phase-matched. The sample transmittance at the pump and the SHG signals was measured to be approximately 94\% and 89\%, respectively [Fig.~\ref{fig:spectrum}b]. Taking into account these transmission losses in our semi-analytical model, the loss-corrected scaling factor was expected to be $\beta_\mathrm{loss}=1.60$. 

In addition to losses, also the bandwidth of the input pump ($\Delta \lambda_\mathrm{FWHM}=9.5$~nm near 950~nm) affects the SHG power dependence because the sample should be phase-matched for all the wavelengths within the pump bandwidth. As this was not the case with our sample, we also took into account this partial phase mismatching in our model, resulting in a further reduced scaling factor of $\beta_\mathrm{pulsed}=1.25$. Although the used model was simple, e.g.~changes in the pump beam intensity were neglected, the calculated scaling factor value agrees quite well with the measured value of $\beta_\mathrm{exp}=1.12$.

In future, the strength of the SHG signal could be straightforwardly enhanced by further increasing the number of passes $N$ and/or the scaling factor $\beta$. Based on our calculations and experiments, it seems advantageous to focus first on increasing $\beta$. A straightforward approach to increase $\beta$ would be to reduce the transmission losses of the relevant wavelengths by incorporating anti-reflection coatings into the metasurface devices. Furthermore, the scaling factor could be increased (simultaneously with $N$) also by moving from monolayer metasurfaces to using stacked metasurfaces~\cite{Stolt:21}, which would reduce the overall transmission losses by minimizing the number of interfaces.

Once transmission losses are properly remedied, it becomes relevant to tackle the reduction of $\beta$ due to partial phase mismatch when spectrally broad fs-pulses are used as the pump source. One could for example reduce the amount of normal dispersion of the system by reducing the thickness of the used glass substrate, on top of which the metasurfaces have been fabricated. In addition, one could design the metasurfaces to be anomalously dispersive, flattening the overall dispersion profile of the system. Instead of tackling the problems associated with spectrally broad pumps, in future one could also focus on nonlinear applications where spectrally narrow pump beams, i.e.~continuous-wave lasers, would be used.

We note that it would be challenging to further increase $N$ using our current experimental setup. However, finding ways to increase $N$ would be in future of paramount interest. %in future it would be very relevant to find approaches also to increase $N$.  
The current challenges were attributed to the finite length of the Rayleigh range of the laser beam and to the physical dimensions of the sample. Together, these issues necessitated us to separate the two mirrors by several millimeters. We think that one solution to above would be to effectively increase the Rayleigh range by using slightly curved (convex) mirrors~\cite{Milosevic:00,Schulte:16,Goncharov:23}.

As other potential approaches to further increase $N$, we are also considering to fabricate metasurfaces directly on top of the mirrors, or completely replacing the mirrors with  highly-reflecting metasurfaces~\cite{deVos:19,Geromel:23}. This replacement would allow the metasurfaces to serve both as sources of harmonic signals and as substitutes of the mirrors. This way the propagation distance between each pass could be considerably reduced, leading to higher $N$ and improved efficiency. In addition, this approach would reduce the amount of normal dispersion, simultaneously increasing the achievable $\beta$, sounding therefore as an effective approach to considerably increase achievable conversion efficiencies.

\section{Conclusions}
 In this study, we have demonstrated an order of magnitude enhancement of the SHG signal in nonlinear metasurfaces by utilizing a simple multipass configuration, where the pump beam traverses the metasurface multiple times. We designed and fabricated a nonlinear metasurface composed of random array of V-shaped aluminum nanoparticles. In our multipass configuration, the pump beam undergoes up to 9 passes through the sample, resulting in an enhancement of the SHG signal at specific wavelengths that satisfy the phase-matching condition.
Our findings, validated by the agreement between our semi-analytical model and experimental observations, underscore the robustness of our approach. 
Notably, this study introduces an innovative strategy for achieving phase matching of the SHG signal within nonlinear metamaterials, eliminating the need for complex fabrication procedures required to realize phase-matched stacked metasurfaces. Furthermore, our work demonstrates a clear pathway to systematically increase the efficiencies of metasurface-based nonlinear devices.

\section*{Funding}
We acknowledge the support of the Academy of Finland (Grant No. 308596), the Flagship of Photonics Research and Innovation (PREIN) funded by the Academy of Finland (Grants No. 320165 and 320166). 
TS acknowledges also Jenny and Arttu Wihuri Foundation for doctoral research grant. TKH acknowledges Academy of Finland project number (322002).
RF  acknowledges the support of the Academy of Finland through the Academy Research Fellowship (332399).

\section*{Supporting information}
Supporting Information: Details regarding sample fabrication, transmission setup, SHG signal detection, calculations of second harmonic signal generation.

%\section*{Acknowledgements}

\bibliography{references}

\end{document}